\documentclass{aal}
\usepackage{graphicx}
\usepackage{enumerate}
\usepackage{txfonts}
\usepackage{natbib}
\newcommand{\msun}{\ensuremath{\mathrm{M}_\odot}}

\bibpunct{(}{)}{;}{a}{}{,}
\usepackage{color}

\begin{document}

\title{The role of neutron star mergers \\ in the chemical evolution of the Galactic halo}

\author {G. Cescutti\inst{1} \thanks {email to: cescutti@aip.de},
  D.~Romano\inst{2}, F.~Matteucci\inst{3,4,5}, C. Chiappini \inst{1},
  \and R. Hirschi \inst{6,7}} \institute{Leibniz-Institut f\"ur
  Astrophysik Potsdam, An der
  Sternwarte 16, 14482, Potsdam, Germany \\
  \and
  INAF, Osservatorio Astronomico di Bologna, Via Ranzani 1, I-40127 Bologna, Italy \\
  \and
  Dipartimento di Fisica, Sezione di Astronomia, Universit\`a di Trieste, Via G.~B. Tiepolo 11, I-34143 Trieste, Italy \\
  \and
  INAF, Osservatorio Astronomico di Trieste, Via G.~B. Tiepolo 11, I-34143 Trieste, Italy \\
  \and
  INFN, Sezione di Trieste, Via A.~Valerio 2, I-34127 Trieste, Italy  \\
\and Astrophysics Group, Keele University, Keele ST5  5BG, UK; \\
\and Kavli Institute for the Physics and Mathematics of the
  Universe (WPI), University of Tokyo, Kashiwa 277-8583, Japan}

\date{Received xxxx / Accepted xxxx}

\abstract {The dominant astrophysical production site of the r-process
  elements has not yet been unambiguously identified. The suggested
  main r-process sites are core-collapse supernovae and merging
  neutron stars.} 
{We explore the problem of the production site of Eu.  We also 
  use the information present in the observed
  spread in the Eu abundances in the early Galaxy, and not only its
  average trend. Moreover, we extend our investigations to  other heavy elements 
(Ba, Sr, Rb, Zr) to provide additional constraints on our results. }
{We adopt a stochastic chemical evolution model that takes
  inhomogeneous mixing into account. The adopted yields of Eu from merging neutron stars
  and from core-collapse supernovae are those that 
 are able to
 explain the average [Eu/Fe]--[Fe/H] trend observed for solar
  neighbourhood stars, the solar abundance of Eu, and the present-day
  abundance gradient of Eu along the Galactic disc in the framework of
  a well-tested homogeneous model for the chemical evolution of the
  Milky Way. Rb, Sr, Zr, and Ba are produced by both the s- and
  r-processes. The r-process yields were obtained by scaling the Eu
  yields described above according to the abundance ratios observed in
  r-process rich stars.  The s-process contribution by spinstars is
  the same as in our previous papers.}  
{  Neutron star binaries that merge in less than 10~Myr or neutron star
  mergers combined with a source of r-process generated by massive
  stars can explain the spread of [Eu/Fe] in the Galactic halo. The
  combination of r-process production by neutron star mergers and
  s-process production by spinstars is able to reproduce the
  available observational data for Sr, Zr, and Ba. We also show the
  first predictions for Rb in the Galactic halo. }
{We confirm previous results that either neutron star mergers on a
  very short timescale or both neutron star mergers and at least a
  fraction of Type II supernovae have contributed to the
  synthesis of Eu in the Galaxy. 
 The r-process production of Sr, Zr, and Ba by neutron star mergers
  - complemented by an s-process production by spinstars - provide results that are
compatible with  our previous 
findings based on other r-process sites.
We critically discuss the weak and strong points of both neutron star
merging and supernova scenarios for producing Eu and eventually
suggest that the best solution is probably a mixed one in which both
sources produce Eu. In fact, this scenario reproduces the
scatter observed in all the studied elements better.}


\titlerunning{The role of NSM in the CE of the halo }

\authorrunning{Cescutti et al.}

\maketitle

\section{Introduction}

The heavy element Eu is an {\it \emph{r-process element}}; that is to say, it is
produced by neutron captures on heavy elements (for instance as Fe) in
a rapid process, where rapid refers to the timescale of 
the neutron capture rates relative to
the
$\beta$-decay rates of unstable nuclei. The main production site of Eu
is still a matter of debate \citep[e.\,g.]{Thielemann11}

Observations of heavy element abundances in Galactic halo stars
provide important constraints on the astrophysical site(s) of
r-process nucleosynthesis. Interestingly, a wide spread is found in
the [Eu/Fe] ratios in halo stars (as well as for several s-process
elements, such as Ba, Y, La), and is much wider than the spread found
for [$\alpha$/Fe] ratios in the same stars.

Using an homogeneous chemical evolution model, \citet{Mennekens14} conclude
that neutron star/black hole mergers could be responsible for the Galactic
r-process production. Their model did not consider a further
contribution from Type II supernovae (SNeII).  More recently,
\citet{Matteucci2014} have employed a detailed chemical evolution
model \citep{RKT10} to study the evolution of Eu in the Galaxy.  Two
possibilities for Eu production were considered: i) production by
core-collapse SNe (stars with initial masses from 9 to 50 $M_{\odot}$)
during the explosion, and ii) neutron star mergers (NSM).  The
classical production site for r-process elements, hence Eu, is
core-collapse SNe \citep{Truran81,Cowan1991}.

The reason for
introducing neutron star mergers as an alternative to Eu production
resides in the large uncertainties present in hydrodynamical
nucleosynthesis calculations for r-process elements in massive stars,
in particular, that neutrino winds in SNII explosions are
proton-rich or only slightly neutron-rich \citep[see for example][and
references therein] {AJS07,WJM11,ArconesThielemann13} and therefore
have difficulty producing Eu. At the present time, only the
magneto-rotational driven SNe (MRD SNe) scenario has been shown to be
a promising source of the r-process by \citet{Winteler12} in the context
of massive stars, and this result has been confirmed very recently by
\citet{Nishimura15}.  However, given the specific
configuration needed by these progenitors, they are expected to be
rare. On the other hand, the nucleosynthesis calculations relative to
neutron star mergers have provided robust results concerning the
r-process element production in these objects \citep[see for
example][]{Rosswog99,Rosswog00, Oechslin07, Bauswein13, Rosswog13,
  Hoto13, Kyutoku13}. 

It has been suggested that up to
10$^{-2}$~M$_\odot$ of r-process matter may be ejected in a single
coalescence event.  \citet{Matteucci2014} suggest that neutron star
mergers could be entirely responsible for the Eu production in the
Galaxy if the coalescence timescale is no longer than 1 Myr for the
bulk of neutron star binary systems, the average Eu yield is
5~$\cdot$ 10$^{-6}$~M$_{\odot}$, and the mass range of progenitors of
neutron stars is 9--50~M$_{\odot}$. They also conclude that a mixed
scenario could be acceptable, where both merging neutron stars and
core-collapse SNe 
contribute to the Eu production.
In the mixed scenario, the Eu yields
from merging neutron stars should be lower since 
core-collapse SNe
contribute to the enrichment. In particular, it was concluded that SNe
in the range 20--50~M$_{\odot}$ should produce
10$^{-7}$--10$^{-8}$~M$_{\odot}$ of Eu each. Both models could reproduce
the average trend of [Eu/Fe] versus [Fe/H] in the solar neighbourhood,
the solar Eu abundance, and the Eu abundance gradient along the
Galactic disk.

By relaxing the instantaneous mixing approximation, it is also
possible to explore the information contained in the observed scatter
(or lack of) in the different abundance ratios and, in particular, in
[Eu/Fe]. \citet{Argast04} explored the impact of Eu production by
merging neutron stars and SNeII
In their model the diffusion of the
stellar ejecta into the interstellar medium is treated dynamically,
hence predicting the chemical spread in the chemistry of the
surrounding gas. Unlike the approach taken in
\citet{Matteucci2014}, \citet{Argast04} do not provide a model where
both the NSM and core-collapse SN are
simultaneously taken into account. Another approach was presented by \citet{Cesc08} who
modelled the inhomogeneous mixing of the interstellar medium by means
of a stochastic chemical evolution model. However, in this case the
author considered only SNeII as a site of production of the r-process.
Both in \citet{Argast04} and \citet{Cesc08}, a large abundance scatter
is predicted for neutron capture elements, whereas a  much lower scatter
is found for alpha/Fe abundance ratios, in agreement with
observations.  In both cases this was interpreted as a
consequence of the stochastic formation of massive stars coupled with
the different stellar mass ranges from which different elements come. In particular, \citet{Cesc08} suggested that the wide spread
observed in neutron capture elements and the  significantly narrower spread in
$\alpha$-elements occurs because the site of production of
$\alpha$-elements includes the whole range of massive stars from 10 to 80
M$_\odot$ whereas the mass range of production for neutron capture
elements lies between 12 and 30 M$_\odot$.   More recently,
  cosmological SPH simulations that include treating the chemical
  elements have investigated the NSMs as possible sources of the r-process
\citep{vandenVoort15,Shen14}.

It is now important to check for consistency between the results
obtained by \citet{Matteucci2014} on Eu with other r-process elements as
well. In particular, it has recently been shown that
magneto-rotationally driven 
(MRD)
supernovae \citep{Winteler12}
represent a promising source of r-process in the early Galaxy
\citep{Cescutti14}. This model was able to reproduce the
observed spread in the abundance ratios not only of Eu, but also of Sr,
Ba, and Y. In the case of Sr, Ba and Y parts of the production most
likely came from spinstars \citep{Pigna08, Frisch12}, and the spinstar contribution to Eu
is expected to be negligible. It has been shown
that including the contribution of spinstars plays a key role in
explaining the long-standing problem of the observed scatter in
[Sr/Ba] in the Galactic halo, as first suggested in
\citet{Chiappini11} and later demonstrated by the inhomogeneous model
calculations of \citet{Cescutti13}. Interestingly, the spinstar
scenario also plays a key role in explaining light element
observations such as C and N
\citep[see][]{Chiappini06,Chiappini08,Cesc10}.

The goal of the present work is to evaluate the impact of including
a neutron star merging scenario (which produces Eu, but also the other
n-capture elements mentioned above) on the previous conclusions based
on the MRD SNe plus spinstar scenario. In the present work we again provide
our predictions for Ba and Sr but also include two other light
neutron-capture elements that have not been modelled before, Zr and Rb. In
particular, Rb  has only been measured in a few globular
clusters \citep[][]{Barbuy2014,Yong14,Dorazi13,Yong08,Wallerstein07}.

The paper is organized as follows. In Section 2 we summarize the observational data considered in this work. In Section 3, we introduce the chemical evolution model. In Section 4
we present our results and compare them to the available observations. In
Section 5, we draw some conclusions.

\section{Observational data}

We employed the same data as used in \citet{Cescutti13}: the data
compiled by
\citet{Frebel10}\footnote{http://cdsarc.u-strasbg.fr/cgi-bin/qcat?J/AN/331/474} and
labelled as halo stars\footnote{The list of authors we use from the
  collection are \citet{MCW95}, \citet{MCW98}, \citet{ WES00}, \citet{
    AOK02b}, \citet{ COW02}, \citet{ IVA03}, \citet{ Honda04},
  \citet{AOK05}, \citet{ BAR05}, \citet{ AOK06b}, \citet{ IVA06},
  \citet{ MAS06}, \citet{ PRE06}, \citet{ AOK07c}, \citet{ Franc07},
  \citet{ LAI07}, \citet{COH08}, \citet{ LAI08}, \citet{ ROE08},
  \citet{ BON09}, \citet{HAY09}}.  We excluded all upper limits and
carbon-enhanced, metal-poor (CEMP) stars.  For CEMP stars we adopt the
definition given by \citet{Masseron10}, where a CEMP star has 
[C/Fe]$>$0.9.  In the compilation by Frebel, there is also a large
portion of stars without carbon measurements. For these stars we
cannot establish whether they are CEMP stars or not; nevertheless, since they
represent a large portion, we decided to include them in our plots, but
to distinguish them graphically from the confirmed normal stars. 
  CEMP-s stars are excluded from our comparison because, if carbon
  enhancement is caused by transfer of matter from an evolved
  companion, the abundances of s-process elements are likely to be
  affected, too. 
 Therefore, no meaningful comparison can be done with the predictions
  of our chemical evolution model in this case, since it refers to the
  chemical composition of the stars at birth.   

We expect CEMP-no
    stars to behave differently from normal stars only
    in the abundances of their light elements 
 \citep{Cesc10, Maeder14}.
  The abundances of their heavy elements show features compatible to
    those of normal stars \citep{Cescutti13}. Nevertheless, since they
    are not the main focus of this work, CEMP-no stars are not included in the present analysis.
 We did not show any data for Rb, since for this element we could not find
data for field stars at extremely low metallicity, and to our knowledge
the lowest metallicity data for Rb in the literature refer to stars
with [Fe/H] $\sim -$2 that belong to globular clusters measured by
\citet{Yong06,Yong08}.

\section{The chemical evolution model}

The chemical evolution model adopted here is the same as in
\citet{Cescutti13} and \citet{Cescutti14}.  
We review its main characteristics here for the reader's convenience.

We considered the chemical evolution model presented in
\citet{Cesc10}, which is based on the inhomogenous model developed by
\citet{Cesc08} and on the homogeneous model of \cite{Chiappini08}. 
The halo consists of many independent regions, each with
the same typical volume, and each region does not interact with the
others.  Accordingly, the dimensions of the volume are  expected to be large 
enough to allow us to neglect the interactions between different volumes, 
at least as a first approximation. For typical ISM densities, a
supernova remnant becomes indistinguishable from the ISM -- that is,
merges with the ISM -- before reaching $\sim50$ pc \citep{Thornton98};
 therefore, we decided to have a typical volume with a radius of roughly
90 pc, and the number of assumed volumes is 100 to ensure good
statistical results.  We did not use larger volumes because
we would lose the stochasticity we are looking for; 
in fact, larger
volumes produce more homogeneous results.

In each region, we assumed the same law for the infall of the gas with
primordial composition, following the homogeneous model by
\citet{Chiappini08}:

\begin{equation}
\frac{dGas_{in}(t)}{dt} \propto e^{-(t-t_{o})^{2}/\sigma_{o}^{2}}, 
\end{equation}
where $t_{o}$ is set to 100 Myr, and $\sigma_{o}$ is 50 Myr.
Similarly, the star formation rate (SFR) is defined as
\begin{equation}
SFR(t) \propto ({\rho_{gas}(t)})^{1.5},
\end{equation}
where $\rho_{gas}(t)$ is the density of the gas inside the  volume under consideration.
Moreover, the model takes an outflow
from the system into account:
 \begin{equation}
\frac{dGas_{out}(t)}{dt}  \propto SFR(t).
\end{equation}

Knowing the mass that is transformed into stars in a time step
(hereafter, $M_{stars}^{new}$), we assigned the mass to one
star with a random function, weighted according to the initial mass
function (IMF) of \citet{Scalo86} in the range between 0.1 and
100~$M_{\odot}$.  We then extracted the mass of another star and
repeated this cycle until the total mass of newly formed stars
exceeded $M_{stars}^{new}$.  In this way, $M_{stars}^{new}$ is the
same in each region at each time step, but the total number and mass
distribution of the stars are different. We thus know the mass of each
star contained in each region, when it is born and when it will die,
assuming the stellar lifetimes of \citet{MM89}.  At the end of its
lifetime, each star enriches the ISM with its newly produced chemical
elements and with the elements locked in that star when it was formed,
excluding the fractions of the elements that are permanently locked
in to the remnant.

As shown in \citet{Cescutti13}, our model is able to reproduce the MDF
measured for the halo by \citet{Li10}. This comparison shows that the timescale of
enrichment of the model is compatible with  that of the halo stars
in the solar vicinity. Moreover, our model predicts a small spread for
the $\alpha$-elements Ca and Si, 
which is compatible with the observational data.

\subsection{Stellar yields for Eu}
\label{sec:yields}
For the Eu production sites, we consider NSMs
and core-collapse SNe, as mentioned in the Introduction.  
 To take the Eu production from
NSMs into account, we need to define the following quantities \citep[see][]{Matteucci2014}:
   \begin{enumerate}
   \item the fraction of massive stars belonging to double NS systems that will eventually merge,
namely the realization probability for such events, $\alpha_{\mathrm{NS}}$;
   \item the time delay between the formation of the double neutron star system 
     and the merging event, $\Delta t_{\mathrm{NS}}$;
   \item the amount of Eu produced during the merging event, 
     $M^{\mathrm{Eu}}_{\mathrm{NS}}$.
   \end{enumerate}
Concerning NSM yields, we also follow what
is assumed in \citet{Matteucci2014};
 in particular, we assume the yields of \citet{Korobkin12}, who suggest
that NSM can produce from $10^{-7}$ to $10^{-5}$ M$_{\odot}$ of Eu per event.

   In the present paper, we assume that a fixed fraction of all the
   massive stars formed in our simulation is a progenitor of NSMs
   and produces r-process material.  The progenitors are
   randomly chosen among the massive stars formed in the range 9-30
   M$_{\odot}$.  The prescriptions for the different models are
   summarized in Table \ref{tabmodels}, where we list (i) the model
   name (column 1), (ii) the delay time for coalescence of NS in
   binary systems (column 2), (iii) the fraction of massive stars that
   are hosted in binary systems leading to NSMs (column 3); (iv)
   the mass ejected as newly produced Eu in NSM events (column
   4), and (v) the Eu yields from massive stars.  For Models NS03
   and NS04, the amount of r-process in the single event is not 
     constant, but we considered the possibility that the
     amount of mass ejected as r-process varies. Since the variation
     is unknown, we assumed the following range: 
the minimum production is 1\% of the average Eu amount, and the maximum is twice the average.
     Since
     the total production should be conserved, the ejected mass for the $n$-th
     star (r-process producer) in these models can be described by the
     following equation:

\begin{equation}
M^{\mathrm{Eu}}_{\mathrm{NS}}(n) = M^{\mathrm{Eu}}_{\mathrm{0}}(0.01+1.98\cdot Rand(n)), 
\label{eqran}
\end{equation}
where $Rand(n)$ is a uniform random distribution in the range
[0,1]. This function is the same
as the one  presented in \citet{Cescutti14}. 
In Table \ref{tabmodels} we report the average Eu amount for the merging event
($M^{\mathrm{Eu}}_{\mathrm{0}}$) assumed for each model.

For the time delay due to the coalescence of the two NSs, $\Delta
t_{\mathrm{NS}}$, \citet{Argast04} and \citet{Matteucci2014} 
considered different timescales: 1~Myr, 10~Myr, and 100~Myr. Here we consider the same timescales. It is worth noting that in previous
works, as well as in this one, it is assumed that all neutron star
binaries have the same coalescence timescale. Clearly, a more
realistic approach should consider a distribution function of such
timescales, in analogy with SNeIa for which a distribution for the
explosion times is defined \citep[see][]{Greggio05}.

Among core-collapse SNe, different candidates for Eu production have been
studied in the past, and they can either have low mass (8--10~M$_{\odot}$) or high mass
progenitors \citep[$>$~20~M$_{\odot}$][]{Cowan1991, WWM94,
 Ishimaru99,Trava99,Wanajo2001,Argast04,
 Cesc06}. The yields of Eu from SNe~II that we adopt here are similar to those of
\citet{Argast04} (their model SN2050) modified as in
\citet{Matteucci2014} (their model Mod2SNNS), as shown in Table
\ref{tabmodels}. They are coupled with a production from NSM with a
delay of 1~Myr (see Table \ref{tabmodels}).  Actually, since we adopt
iron yields from SNeII larger than \citet{Matteucci2014}, we need to
slightly increase the Eu production for the 50~$\msun$ to balance the
higher production of iron in our models.  Since
the most recent results concerning the production of Eu in SNe do not
confirm these channels (see Arcones et al. 2007, for the high mass
channel, and Wanajo et al. 2011, for the low mass channel), we investigate
another possible production connected to massive stars: the
magneto-rotational driven (MRD) scenario.  \citet{Winteler12} have shown
that the combination of high rotation and strong magnetic field in the
inner core of an exploding SN promotes an r-process production \citep[see also][]{Nishimura15}. This
specific configuration is rare, therefore only a limited number of
SNe have this fate, and as mentioned in \citet{Winteler12},
it should be more frequent in the metal-poor regime \citep{Yoon06}.  Therefore, we
explore a model in which 10\% of the SNe produce r-process material,
with the same prescriptions as assumed in \citet{Cescutti14}.  However,
unlike that paper, we consider here that this channel is
only active at low metallicity, $Z<10^{-3}$. Coupled with the production of
MRD SNe, we also assume a NSM production with a fixed delay of 100~Myr
(see Table \ref{tabmodels}).

\begin{table*}
\begin{center}
\caption{Prescriptions for Eu production for different models. See text for explanations.
}\label{tabmodels}
\begin{tabular}{|c|c|c|c|c|}
 \hline
  Model name&  $\Delta t_{\mathrm{NS}}$ & $\alpha_{\mathrm{NS}}$ &M$^{Eu}_{0}(M_{\odot})$ &  M$^{Eu}_{newly produced}(M_{\odot})$\\  
                   &                    &   &  neutron star mergers   &  massive stars  \\  
\hline
  NS00 & 1 Myr & 0.02 & 5$\cdot10^{-6}$ (constant per merging event)& no production\\
\hline
  NS01&   10 Myr &  0.02 & 5$\cdot10^{-6}$  (constant per merging event)& no production \\
\hline
  NS02 & 100 Myr &  0.02 & 5$\cdot10^{-6}$ (constant per merging event)& no production\\
\hline
  NS03 & 1 Myr & 0.02 &on average 5$\cdot10^{-6}$ (varying as in equation~\ref{eqran})& no production\\
\hline
NS04 & 1 Myr & 0.04 &on average 2.5$\cdot10^{-6}$ (varying as in equation~\ref{eqran})& no production\\
\hline
NS+SN & 1 Myr & 0.02 & 3$\cdot10^{-6}$ (constant per merging event)&  2$\cdot$ 10$^{-8}$--5$\cdot$10$^{-7}$\\
& & & & (linear interp. in the range 20-50 M$_\odot$)\\
\hline
NS+MRD & 100 Myr & 0.02 & 1.5$\cdot10^{-6}$ (constant per merging event) &  on average 1$\cdot10^{-6}$  for 10\% of stars in 8-80M$_\odot$\\
& & & &  (varying as in equation~\ref{eqran})\\

\hline 
 \end{tabular}
\end{center}

\end{table*}  

Finally, we note that in the spinstar framework, Eu is produced in negligible amounts.

\subsection{Stellar yields for Rb, Sr, Zr, and Ba}

Rb, Sr, Zr, and Ba are produced by both the s- and
r-processes.
The r-process yields are obtained by scaling the Eu yields adopted
here according to the abundance ratios observed in r-process-rich
stars \citep{Sneden08}. Another possible choice would be to take the solar system r-process contribution into
account as determined, for
example, by \citet{SSC04} and \citet{Arla99}; we checked that this does not
seriously affect the results for these elements.   Our choice does
  not rest on the results of the theoretical computations of the main
  r-process; rather, we infer the r-process elements yields from an
  observational signature that could reflect a combination of
  processes (e.\,g. main r-process + weak r-process). Moreover, a given
  process might present some intrinsic variation. Indeed, from an
  observational point of view, it seems that the robust pattern for
  the r-process does not extend to the elements between the first and
  second r-process peaks even within the class of highly r-process-enhanced stars \citep{Roederer14}.  In the future, we plan to use
  theoretical results for the r-process ratio to investigate this
  aspect.

The spinstars' s-process contribution for all our models is the same as
in the \emph{fs-}model of \citet{Cescutti13}. However, we show here
results for rubidium and zirconium, which were not treated in the
previous paper, therefore we recall that for the yields at Z
=$10^{-5}$, we considered the stellar yields obtained by
\citet{Frisch12} after decreasing the reaction rate for
$^{17}O(\alpha,\gamma)$ from \citet{CF88} by a factor of 10, which
enhances the s-process production\footnote{A value of the metallicity of Z=$10^{-5}$ corresponds to [Fe/H]$\simeq
  -$3.5, with small variations due to the stochasticity of the
  models.}.  Unfortunately, we only have results
with this reaction rate for a single mass (25~\msun) at
Z=$10^{-5}$, and we used the scaling factor obtained for the whole
range of masses \citep[for more details, see][]{Cescutti13}. Indeed,
there are no nucleosynthesis calculations for spinstars currently
carried out with a reduced value of the $^{17}O(\alpha,\gamma)$ rate
for a metallicity higher than Z=$10^{-5}$, and we adopted those
computed with the standard value given by \citet{CF88}. We need to
keep this caveat in mind when interpreting our theoretical predictions
for the intermediate metallicity range.  We also considered the
s-process contribution from stars in the mass range 1.3-3$M_{\odot}$, by implementing the yields by
\citet{Cristallo09,Cristallo11} in the models. We underline, however, that this production channel affects the model results only at moderate
metallicity ([Fe/H]$\sim$-1.5).

\section{Results}
\subsection{NSM models}

\begin{figure*}[ht!]
\begin{minipage}{180mm}
\includegraphics[width=180mm]{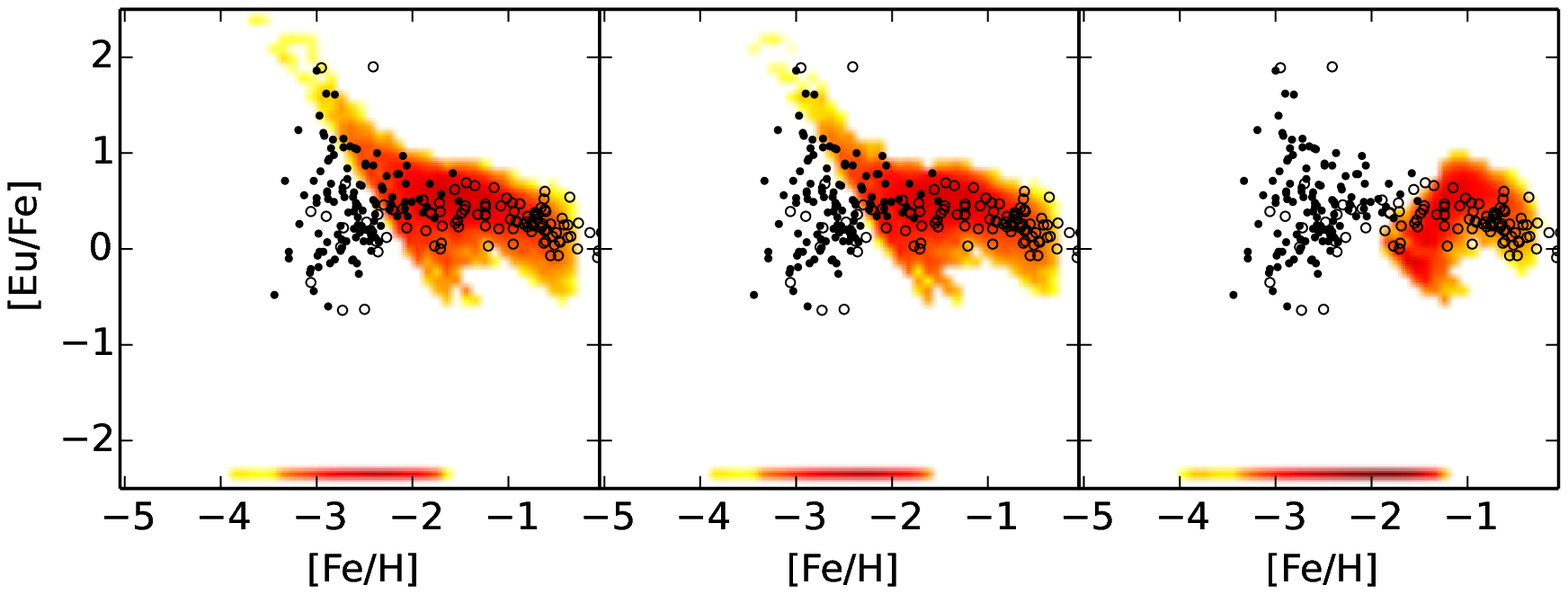}
\begin{center}
\vspace{-10mm}
\includegraphics[width=120mm]{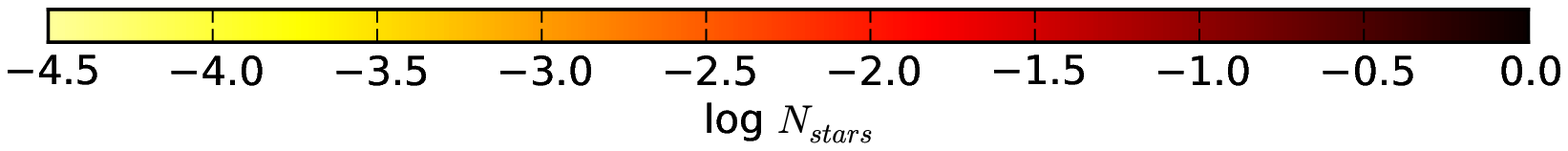}
\caption{Left panel, the results for [Eu/Fe] vs [Fe/H] for model
  NS00. This model has a delay time for the neutron star mergers of
  1~Myr, constant Eu production of 5$\cdot10^{-6}$ M$_{\odot}$ per
  merging event, and no Eu from SNeII.  The density plot is the
  distribution of simulated long-living stars for our model (see the
  bar below the figure for the colour scale). The long-living stars
  formed without Eu (formally [Eu/Fe]=$-\infty$) are
  shown at [Eu/Fe]=$-$2.4.  The model predictions are compared to data
  collected in \citet{Frebel10}; we show as black dots stars with
  [C/Fe] $<$0.9 (to avoid binary enrichment), the open dots are stars
  with no carbon measurement.  Central panel, same as left panel, but
  for model NS01. This model has a delay time for the neutron star
  mergers of 10~Myr, constant Eu production of 5$\cdot10^{-6}$
  M$_{\odot}$ per merging event, and no Eu from SNeII. Right panel,
  again same as for the left panel, but for model NS02. This model has a delay
  time for the neutron star mergers of 100~Myr, constant Eu production
  of 5$\cdot10^{-6}$ M$_{\odot}$ per merging event, and no Eu from
  SNeII. }\label{figA}
\end{center}
\end{minipage}
\end{figure*}

\begin{figure*}[ht!]
\begin{minipage}{185mm}
\includegraphics[width=185mm]{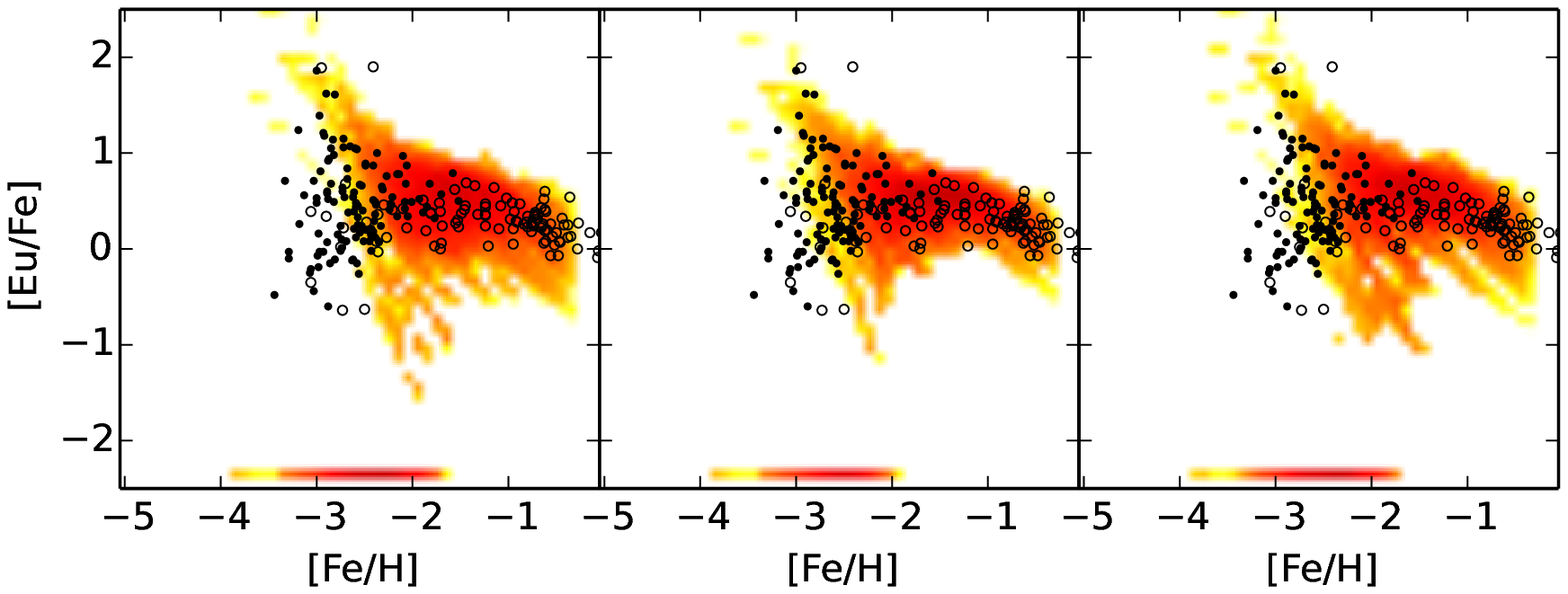}
\vspace{-10mm}
\caption{ Left panel, same as Fig.~\ref{figA}, but for model
  NS03. This model has a delay time for the neutron star mergers of 1~Myr,
  variable Eu yield per merging event (average value 5$\cdot$10$^{-6}$
  M$_{\odot}$), and no Eu from SNeII. Central panel, same as left panel, but for
  model NS04. This model has a delay time for the neutron star mergers of 1~Myr,
  variable Eu yield per merging event (average value
  2.5$\cdot$10$^{-6}$ M$_{\odot}$), and no Eu from SNeII. In this
  model, the fraction of massive stars leading to NS merging events is
  twice what is assumed in previous models. Right panel, same as left panel, but we
  present here a model similar to model NS03 in which we consider
  a mass range for neutron stars progenitors of the NSM of
  9-50~M$_{\odot}$ instead of 9-30~M$_{\odot}$.}\label{figB}

\end{minipage}
\end{figure*}

In Fig.\ref{figA} (left panel) we show the distribution of synthetic halo stars in
the [Eu/Fe]--[Fe/H] plane as predicted by our inhomogeneous model when
assuming that (i) Eu is produced only in NSMs  with progenitors for neutron stars in the range 9-30$M_{\odot,}$; (ii) 2\% of
massive stars are in binary systems with the right characteristics to
lead to merging NS, (iii) each merging event produces
5$\cdot10^{-6}$ M$_{\odot}$ of Eu, and (iv) there is a  fixed delay time of 1
Myr between the formation of the double NS system and the merging
event (model NS00, see Table \ref{tabmodels}).

It is seen that, while explaining very well the data of stars with
[Fe/H]$\ge -$2.2 dex, such a model fails to explain the presence of
stars with [Eu/Fe]$<$0.5 for [Fe/H]$\le -$2.5 and does not explain the
existence of any star with Eu measurements at metallicities lower than
[Fe/H]=$-$3.  The upturn in [Eu/Fe], visible at low metallicities, is
a consequence of the fixed amount of Eu produced by NSM, coupled with
the paucity of NSM events and the constant mixing volume assumed in
our model.  When a NSM pollutes a simulated box early on, it produces
a value in the [Eu/Fe] vs [Fe/H] space, dependent on the mass of the
previous enriching SNeII.  The volume enriched by NSM and SNII with
the lowest amount of iron creates the upper tip of this upturn towards
low metallicity.  Then in all the volumes polluted by NSM, the
probability of having another Eu enrichment is low, so they evolve
towards lower [Eu/Fe] and higher [Fe/H] by the subsequent enrichment
of Fe by SNII, creating the diagonal shape from high [Eu/Fe] with low
[Fe/H] to low [Eu/Fe] with higher [Fe/H] shown
in Fig.\ref{figA} (left panel).  A more detailed dynamical treatment,
where the pollution in not confined to fixed volumes, does not produce
this sharp feature \citep{Argast04}. Also to consider a variation in
the r-process production (as in the models NS03-NS04, see below) leads
to a smoothing of this sharp trend. The fraction of Eu-free stars,
shown in the band at [Eu/Fe]=-2.4, is not negligible for this model
and also for the next NSM models; we discuss the implication of this
outcome in the Sect. 4.4 in more detail.

A model with a delay time of 10 Myr (model NS01) essentially
behaves the same (see Fig.~\ref{figA}, central panel), while delay times
as long as 100 Myr cannot be  accepted, because they lead to even worse
predictions, with all stars with Eu measurements below [Fe/H]$\sim -$2
dex being unexplained by the model (model NS02,
Fig.~\ref{figA}, right panel). Therefore, in the following we only discuss models that
assume $\Delta t_{\mathrm{NS}}$= 1~Myr (apart for model NS+MRD). 
 Similar results have been found by Matteucci et al. (2014).

In Fig.~\ref{figB} (left panel) we show
the predictions of model NS03, which is the same as model NS00, except that the masses ejected by NSMs are not constant, but
instead randomly distributed around the mean value, 5$\cdot10^{-6}$ $\msun$
(see equation~\ref{eqran}). This approach is similar to the one
adopted by \citet{Cescutti14} to explain the Ba, Y, Sr, and Eu
abundance scatter in halo stars in the framework of a model where MRD
SNe are the only r-process element sources. In Fig.~\ref{figB} (central panel) we show
the predictions of model NS04, which differs from Model NS03 in the
adopted value of the average Eu yield from NSMs, $M^{\mathrm{Eu}}_0$=
2.5$\cdot10^{-6}$ M$_{\odot}$, i.e. half the value adopted in Model
NS03. Moreover, Model NS04 assumes that the fraction of massive stars
leading to NSMs is twice the value adopted in Model NS03, namely
4\%. It is worth noticing that \citet{Matteucci2014} require that 2\% of massive stars are hosted in binary systems, leading
to NS mergers in order to fit the current merger rate of these systems
in the Galaxy. However, there are 
no strong arguments against a
variation in this quantity in the early Universe. As a matter of fact,
Model NS04 leads to theoretical predictions that agree with the data
much more closely than Model NS03. In particular, Model NS03 also predicts a
wide spread at intermediate metallicities, which also implies a
non-negligible population of stars with sub-solar [Eu/Fe] ratios at
[Fe/H]$\sim -$2, which are actually not observed.  We note that,
however, Model NS04 does not allow fully addressing the existence
of stars with [Eu/Fe]$<$0 for [Fe/H]$< -$3.  In Fig.\ref{figB} (right panel), we
show the results of Model NS03, but with a mass range for neutron star
progenitors of 9-50~M$_{\odot}$, instead of 9-30~M$_{\odot}$.  In
fact, \citet{Matteucci2014} suggest that Eu in the Galaxy can only be
reproduced by NSM if the range of progenitors of neutron stars
extends up to 50~$M_{\odot}$. However, this change does not provide
significant variations in our predictions, mostly because of
the stochastic nature of our model in which the
enrichment at the lowest [Fe/H] is not necessarily due to the most massive stars.

We are aware that to obtain a reasonable agreement with the
observational data, we had to assume the shortest timescale for the
neutron stars mergers suggested in the literature \citep{Bel2002}, and
this can be an extreme assumption. In fact, a progenitor lifetime of 1
Myr is already the shortest possible timescale in the distribution of
merger times by \citet{Bel2002}; then, unfortunately, their population
synthesis method is not very predictive of the merging timescale:
``Unlike all other binary properties ... the qualitative
characteristics of the merger-time distributions appear to be rather
sensitive to a number of model parameters.''  Moreover, the local rate
of short-hard gamma ray bursts can provide at least a rough
observational constraint on the typical lifetimes of its progenitors
(so NS-NS binaries), and the results seem to indicate lifetimes of a
few Gyrs \citep[see][]{Nakar2006}. Finally, the estimated merger times
for the few NS-NS binaries in our Galaxy are of the order of 100~Myr
\citep{Kalogera2001,Piran2005}, so all these observational indications
cannot exclude that some NSM explode on a very short timescale, but at
least they seem to point to an average merger time that is much longer
than 1~Myr.

\citet{Argast04} have also studied the NSMs as a possible source of
r-process material. Their model dynamically traces the diffusion of
the ejecta, so in this respect it is more detailed than our
model. However, it is not clear how well their model follows the
chemical evolution of the Galaxy, since no comparison with the global
MDF of the halo was provided, and no outflow from the
system was considered.  Still, despite these differences, the results
do not differ substantially, apart 
from the tendency of their model to
produce too many r-process-poor stars at intermediate metallicity
compared to our results and presumably a larger number of metal free
stars.  Also their conclusions are similar: ``NSM as major r-process
sources are only consistent with observations'' if `` the NSM
population has coalescence timescales shorter than approximately 10
Myr.''  They underline that NSM could be co-producers of the r-process
,together with core-collapse SN, but they do not analyse the possible
outcome. We provide this comparison in the next
section.

\subsection{NSM and core-collapse models}

\begin{figure*}[ht!]
\begin{center}
\begin{minipage}{180mm}
\hspace{25mm}
\includegraphics[width=180mm]{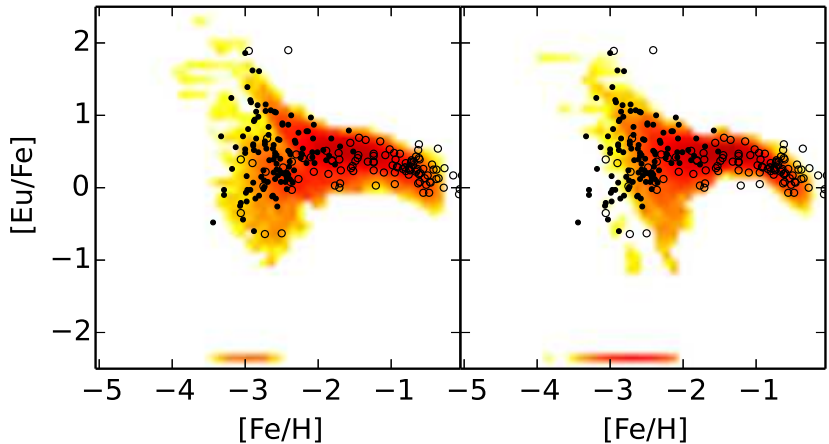}
\vspace{-10mm}
\caption{Left panel: same as Fig.~\ref{figA}, but for Model NS+SN.
  The prescriptions about Eu nucleosynthesis for Model NS+SN are
  similar to those of model Mod2SNNS by \citet{Matteucci2014} , but it
  allows for inhomogeneous mixing in the early Galaxy. Right panel:
  same as left panel but for Model NS+MRD. In this model we consider
  the production of Eu by neutron star mergers with a delay of 100~Myr
  and also a production by MRD SN (10\% of all SNe II) but only for
  $Z<10^{-3}$.}
\label{figC}
\end{minipage}
\end{center}

\end{figure*}

A possible solution to the issue of the existence of stars with
[Eu/Fe]$<$0 at [Fe/H]$< -$3 is to consider Eu production from
core-collapse SNe.  At the present time, a scenario in which all SNe~II produce Eu is not supported by nucleosynthesis models.  A standard
SNII explosion can produce chemical elements only up to the Sr-Y-Zr peak
during a normal core-collapse SN explosion \citep{AJS07}.
Nevertheless, at least to have a possible idea of the impact of this
production, we present the results of model NS+SN in 
Fig.~\ref{figC} (left panel).
This model shares the prescriptions for Eu production
in NSMs with model NS03, but add to it a contribution from SNeII increasing from
2$\cdot10^{-8}$ M$_{\odot}$ for a 20~M$_\odot$ star to 5$\cdot10^{-7}$
M$_{\odot}$ for a 50~M$_\odot$ star. As for the synthesis of Eu, Model
NS+SN adopts prescriptions similar to those of Model Mod2SNNS in
\citet{Matteucci2014}.  The model proved successful in reproducing the
average [Eu/Fe]--[Fe/H] relation of solar neighbourhood stars, the
solar Eu abundance, and the present [Eu/H] gradient across the disc.

Finally, we explore the possibility that a fraction of 
massive stars end their 
life with a strong magnetic field and an
extremely fast rotation in their inner cores.  This MRD SN 
 scenario has been studied in \citet{Winteler12} and \citet{Nishimura15}. According to
their results, the explosion is able to trigger a production of
r-process material.  We assume that this channel is active only
in 10\% of massive stars and only in the metal-poor environments
(log(Z)$<-$3) since this channel is favoured at low metallicities.
  As in previous cases, in Model NS+MRD, the MRD
production is coupled with the NSM productions, but with a long
 delay of 100~Myr.  
In Fig.~\ref{figC} (right panel) the results of Model NS+MRD are presented.  In this
case the model is able to explain the distribution of the data 
at low metallicity and the overall trend very well.

Model NS+MRD provides a possible solution if the timescale for Eu
enrichment from NSM turns out to be relatively long. Clearly,
more detailed assumptions, such as a time delay distribution for the NSM
or a fading of the MRD contribution (rather than a switch off at a
certain metallicity), would be an improvement in the modelling. However,
at the moment neither observational nor theoretical indications allow
us to build better scenarios, so we prefer to keep things simple.
Nevertheless, we underline again that this double scenario is a
promising way to explain the distribution of [Eu/Fe] in extremely
metal-poor stars.

\subsection{NSM and spinstar scenario}

\begin{figure*}[ht!]
\begin{minipage}{185mm}
\begin{center}

\includegraphics[width=185mm]{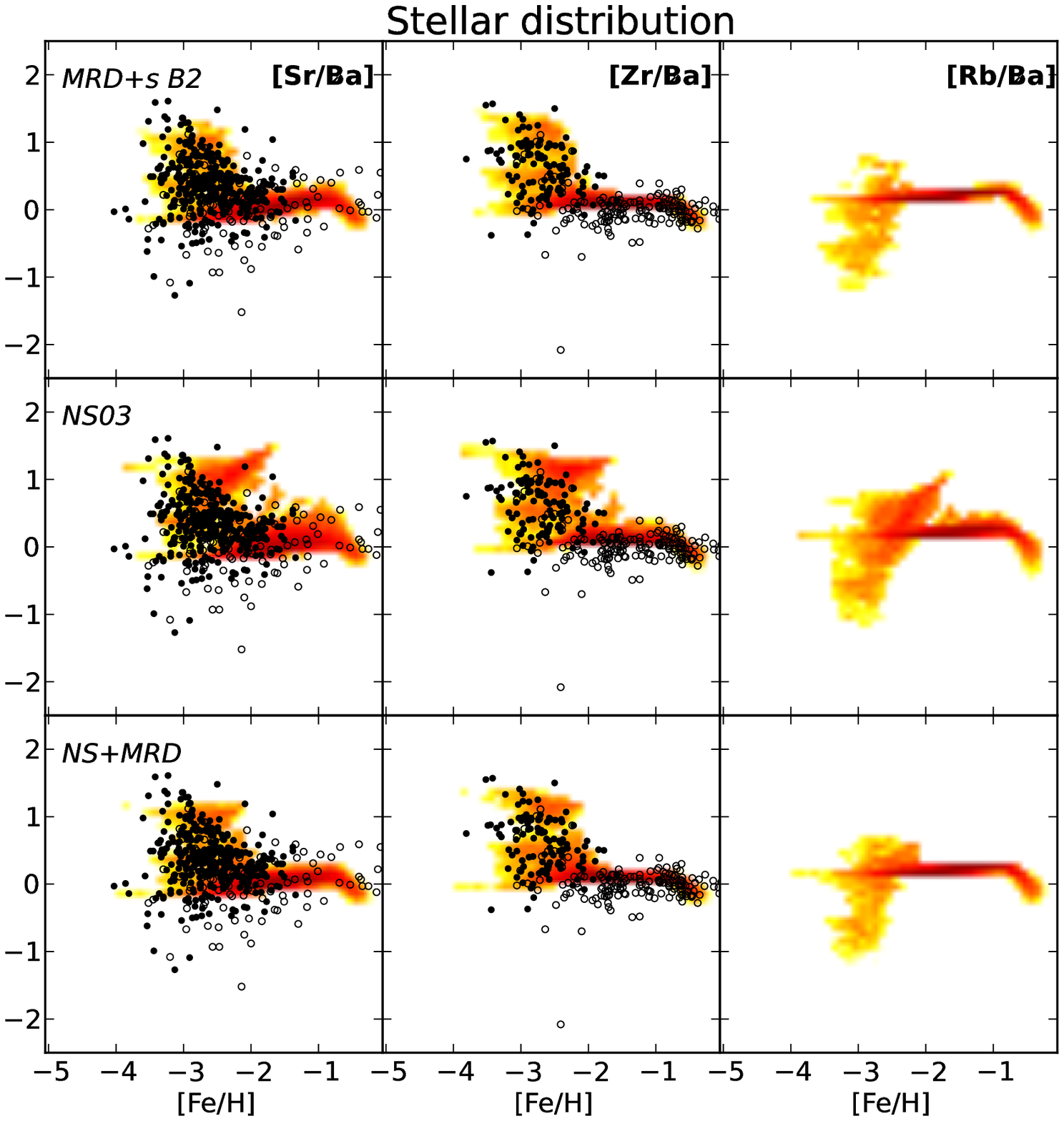}

\caption{From the left [Sr/Ba], [Zr/Ba] and [Rb/Ba] vs [Fe/H] in the
  halo; the density plot is the distribution of simulated long-living
  stars for our models. Superimposed, we show the abundance ratios for halo stars
  \citep[data from][]{Frebel10}. The symbols for the
  \citet{Frebel10} data are black dots for normal stars, 
and black open circles
  for stars without carbon measurement.}\label{extra}

\end{center}
\end{minipage}
\end{figure*}

Now we turn to other neutron-capture elements, with both an r- and an
s-process component, namely Sr, Ba, Zr, and Rb. In the first column of
Fig.~\ref{extra} we show the resulting [Sr/Ba] vs [Fe/H] for Models
NS+MRD and NS03, compared to Model MRD+s B2. 
  In Model  MRD+s B2, we assumed that 10\% of the massive stars end their lives as
MRD SNe, and we considered the
  possibility that the amount of mass ejected as r-process varies
following equation \ref{eqran} \citep[see][for further details]{Cescutti14}.
The aim is to investigate the impact of the
different r-process productions assumed in NS+MRD and NS03 on the
previous results of \citet{Cescutti13} and \citet{Cescutti14}, in
particular on the idea of spinstar production as a viable explanation
for the spread in [Sr/Ba] in the observational data, which is a concept
introduced by \citet{Chiappini11}.
We recall
that in all the models,  Sr, Ba, Zr, and Rb are
produced by both s- and r-processes. The r-process yields are
obtained by just scaling the Eu contribution according to the chemical
abundance ratios observed in r-process-rich stars \citep{Sneden08}.
We consider the contribution by spinstars, following the yields
calculated by \citet{Frisch12}.  The original Model  MRD+s B2 \citep[see][]{Cescutti14} was in fact able
to reproduce the scatter of [Sr/Ba] data at [Fe/H]$<-$2.5, thanks to the
combination of the spinstar production with [Sr/Ba]$>$0 and the
r-process production fixed at [Sr/Ba]$\sim$-0.2.

Model NS+MRD turns out to be very similar to Model MRD+s B2,
and we note just tiny variations in the 
predicted distribution of long-living stars. This happens because the models have
essentially the same nucleosynthesis prescriptions at low metallicity (log(Z) $<-3$).
  At log(Z) $>-3,$ the models are
different: NSMs are the only r-process producers in Model NS+MRD,
but the impact of this change is mitigated by an ISM that is already homogenous and
quite rich in neutron capture elements, so the models do not
show remarkable differences.

On the other hand, an important difference arises in the comparison
between the NS03 model and the other models. Since the NSM
  rate is lower than the assumed rate of MRD SN events, the
  average time needed to have an enrichment of the r-process by NSMs is
  longer in a stochastic realization. Therefore, we find in
  this model volumes that are only polluted by spinstars for a
  longer period, and they can survive without r-process pollution up
  to [Fe/H]$\sim$-2. This can be noticed in the more extended
prediction of long-living stars with high [Sr/Ba] and also in higher
stellar density, in the area [Sr/Ba]$>$0 and $-3<$[Fe/H]$<-2$.

In the second and third columns of Fig.~\ref{extra}, we display our new
results on the impact of spinstars in the chemical evolution of
[Zr/Ba] and [Rb/Ba].  We underline that these chemical predictions for
Rb are the first results available for this element
(probably due to the paucity of observational data). For Zr we have
found few homogeneous models that describe its evolution \citep[among
them][]{Trava04,Cescutti07}.  The model results for [Zr/Ba] are very
similar to the results for [Sr/Ba]; since the observational data
have similar distributions, we can conclude that our models reproduce the stellar abundances well
at low metallicity for
Zr.
Also, the differences amongst the predictions of the different models,
in particular between Model NS03 and Models MRD+s B2 and NS+MRD,
  follow the same pattern for the [Sr/Ba] case.
In contrast, the predictions for [Rb/Ba] are more complex and
show different trends.  The spinstars for these elements do not always
produce a ratio above the r-process signature, but in both
directions (above and below). 

 The different behaviour of Rb with respect to Sr (and Zr) has several
explanations. The first is that the ratio [Rb/Ba] is generally lower than
[Sr/Ba]. This is because Rb is less stable than Sr (and
Ba). This first difference is enhanced in spinstars for the following
reasons. The isotopes produced in spinstars that are most abundant
have different properties for Sr and Rb, in particular their position
along the s-process path and their neutron-richness \citep[see e.\,g.][ and references therein]{Kaeppeler11}. 
For Sr the most abundant isotope produced is $^{88}$Sr, which is the
most neutron-rich stable isotope of Sr. $^{88}$Sr  is a bottleneck of the
s-process path at $N=50,$ and it is one of the main isotopes produced by
the weak s process in massive stars. Its production is generally
high. The situation is different for Rb. This element has only two
stable isotopes, $^{85}$Rb and $^{87}$Rb (as opposed to 4 for Sr). The Rb
isotope most abundantly produced by the weak s process is $^{85}$Rb. The
production of $^{85}$Rb is sensitive to the branching point at $^{85}$Kr. If
neutron densities are high, more $^{85}$Kr capture another neutron instead
of beta-decaying to $^{85}$Rb. This thus reduces the
production of $^{85}$Rb significantly. This reduction is compensated for partially by an
increase in the production of $^{87}$Rb (via beta decay of $^{87}$Kr), but
generally the production of $^{87}$Rb is an order of magnitude lower than
$^{85}$Rb. Neutron densities are high
in the spinstar models considered in this work (with a low O$^{17}(\alpha,\gamma)$
rate). This leads to a generally lower
production of Rb relative to Sr for the reasons listed above and
explains why [Rb/Ba] are lower than [Sr/Ba] (and [Zr/Ba]).

Indeed, the different behaviour of Rb with respect to Sr (and Zr) is
an interesting signature that can be investigated in future
observational campaigns; at the moment, only a few measurements at higher
metallicity in globular clusters are available. Again, the NS03 model
predicts a stronger presence of long-living stars polluted by
spinstars compared to the models MRD+s B2 and NS+MRD.

In summary, the assumptions on the r-process production of the models NS+MRD
and NS03 do not alter the findings of \citet{Cescutti13} and
\citet{Cescutti14}, and the spinstars produce a spread
in [Sr/Ba] and [Zr/Ba] for these models matching the observational data. For [Rb/Ba], no
data are available at the moment to check our results, but we predict
a different spread than for [Sr/Ba] and [Zr/Ba].  The NS03 model
tends to enhance the population of long-living stars polluted by
spinstars
alone, compared to the two other models. This point is not easily
tested, but the tendency does not appear to be displayed at least in
the available observational data \citep[see Fig. 6 in][]{Cescutti14},
since stars with very high [Sr/Ba] ($>0.5$ dex) are not so frequent.

\subsection{Eu- free stars?}

\begin{figure}[ht!]
\begin{minipage}{80mm}
\includegraphics[width=75mm]{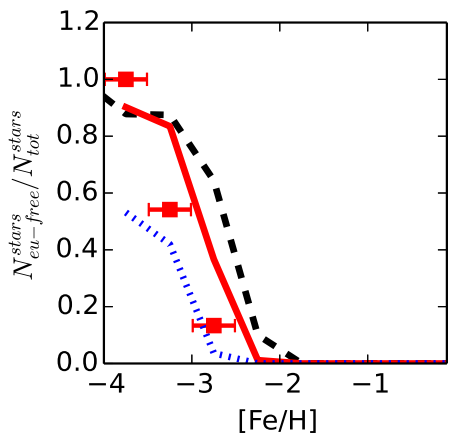}
\caption{Ratio of Eu-free stars over the
  total number of stars for bins of 0.5 dex in [Fe/H]. The model
  results are displayed with a black dashed line (model NS04), red
  solid line (model NS+MRD), and blue dotted line (model NS+SN). The
  red squares are the the observational proxy for this ratio, so the ratio
 between stars in which Eu only presents an upper limit over the number of stars
  for which at least Ba has been measured. The error bars are only plotted
to show the dimension of each bin in [Fe/H].}
\label{fig8}
\end{minipage}
\end{figure}
 The different r-process sites analysed in the previous sections (MRD SN,
NSM) have a common feature: only a small fraction ($<$10\%)
of the massive stars actually enter Eu production.  On one hand, this
promotes the observed spread of [Eu/Fe] at low
metallicities. On the other hand, it ensures that at extremely low metallicities a complementary fraction of low mass stars
are formed in regions where no r-process synthesis took place; for this reason, it seems likely to form some Eu-free stars.
The possibility that the r-process production is delayed further increases the fraction of Eu-free stars, since more stars can be
formed before any Eu enriches the ISM.

Our stochastic models predict the formation of a fraction of Eu-free
stars: all the low mass stars formed in each stochastic realization before the
first enrichment by a r-process site are Eu-free. Therefore, we
investigate how this result matches the observations
of Galactic halo stars.  We check this by computing the ratio of Eu-free stars over the
total number of stars in our models in a given [Fe/H] bin and then
compare these theoretical ratios to an observational proxy for this
ratio. The number of observed stars where it has been measured is only
an upper limit for Eu compared to the total number of stars for which
Ba has been observed 
(see Fig.~\ref{fig8}).  We are aware that there is
probably a bias in this plot, since it is affected by the weakness
of the Eu line, so below a certain metallicity, the S/N to distinguish
the Eu line cannot be achieved during standard observations. For this
reason, we expect this proxy to only be an upper limit of the real fraction
of Eu free stars.

Model NS+SN always shows a lower fraction of
Eu-free stars compared to our observational proxy; since given the above,
this proxy is  an upper limit, it is a positive result.  
The model
NS+MRD predicts a higher number of Eu-free stars than in our
observational proxy. Nevertheless, a slightly higher rate for the MRD
events at extremely low metallicity could solve this problem.
Model NS04, which is the one performing better in the pure NSM scenario, always
predicts a too high ratio of Eu-free stars compared to our
observational proxy, and it also only drops to zero at very high
metallicity. 

We note that also rotating massive stars can produce a tiny amount of
Eu by s-process \citep[see][]{Frisch12}, however the calculations at
the present time can produce at best an [Eu/Fe]$\sim -$4 at
[Fe/H]$\sim -$3. We have indeed included the yields for spinstars in the
models discussed in this paper, but their impact is simply too low to
be seen in the plots.

 Should future observations show that Eu is always present in the
  composition of EMP stars, similar to Ba and Sr (Roederer 2013),
  the most likely explanation would be that all core-collapse SNe are
  producing at least a tiny fraction of Eu through some r-process
  channel (as in our model NS+SN). Another way to solve the riddle,
as  investigated by Komiya et al. (2014), would be to assume that the
  observed Eu is accreted on the stellar surface from the ISM, but
  this result contradicts \citet{Frebel09}

  Overall, the results obtained with the inhomogeneous model for the
  chemical evolution of the Galactic halo adopted in this work confirm
  and reinforce previous conclusions by \citet{Matteucci2014} that
  either NS mergers explode on a very short timescale or that at least
  a fraction of SNeII (MRD SNe) must produce some Eu.  At present, the
  last scenario should be preferred, since it is the one that
  best reproduces the available
  observations.

\section{Conclusions}

In this paper, we adopt the inhomogeneous model for the chemical
evolution of the Galactic halo presented by \citet{Cesc08} and further
developed in \citet{Cesc10,Cescutti14} and \citet{Cescutti13}, to
study the evolution of the n-capture elements Eu, Sr, Ba, Zr, and Rb
in the early Galaxy.  We implemented the Eu production from coalescing
neutron stars in the code following the prescriptions of
\citet{Matteucci2014}. These in turn rest on recent work by
\citet{Korobkin12} showing that large amounts of r-process elements,
0.01 M$_\odot$, can be produced by NS when they merge. Out of these,
the mass of newly produced Eu can lie in the range $\sim$10$^{-7} -
10^{-5}$ M$_\odot$.  \citet{Matteucci2014} have already studied the
effect of such a large Eu
production from NSMs by means of a detailed homogeneous chemical
evolution model for the Milky Way. They suggest that, though NSMs can
in principle explain the history of Eu enrichment in the Milky Way by
themselves, a more realistic situation requires that both NSMs and (at
least) a fraction of core-collapse SNe contribute to the Eu production
in the Milky Way.  However, while \citet{Matteucci2014} could explain
the observed trend of [Eu/Fe] versus [Fe/H] in the solar vicinity, the
abundance of Eu measured in the Sun and the present-day abundance
gradient of Eu along the disc, they did not address the problem of
explaining the wide spread of [Eu/Fe] ratios measured in halo stars,
because their analysis was carried out with the use of homogeneous
chemical evolution models, which are intended to reproduce average
trends.

In the present work we provide a complementary analysis by using
stochastic chemical evolution models and investigate the information
contained in the scatter of several abundance ratios (e.\,g. [Eu/Fe],
[Sr/Ba], [Zr/Ba], and [Rb/Ba]). We show that either NSMs explode with
a very short fixed timescale or both channels -- NSMs and
core-collapse SNe (at least a fraction of them as in the MRD scenario)
-- must be active. In order to explain the spread, as well as both the
presence of stars with low [Eu/Fe] at [Fe/H] $< -$3 and a certain
fraction of Eu free stars, the last scenario should be preferred.

The results shown here are consistent with the recent conclusions by
\citet{Cescutti13} and \citet{Cescutti14} of an important role of
spinstars in the early Universe, not only as contributors to light
elements, but also to n-capture elements such as Sr, Ba, Zr, and
Rb. Indeed, their inclusion, together with NSM and MRD SN sites, leads
to models that can reproduce the scatter observed in all of these
elements very well.

\begin{acknowledgements}
  GC thanks the organizers and participants of the program INT 14-2b
  ``Nucleosynthesis and Chemical Evolution'', where much of this work
  was discussed in Summer 2014; special thanks to Almudena Arcones
  for her intriguing questions and suggestions that have a strong
  impact of this manuscript.  GC also acknowledges financial support
  from the INT during the program; FM and DR acknowledge financial
  support from PRIN MIUR 2010-2011, project 'The Chemical and
  Dynamical Evolution of the Milky Way and Local Group Galaxies',
  prot. 2010LY5N2T. RH acknowledges support from the European Research
  Council (EU-FP7-ERC-2012-St Grant 306901) and from the World Premier
  International Research Center Initiative, NEXT, Japan.

\end{acknowledgements}

\bibliographystyle{aa}
\bibliography{spectro}

\begin{thebibliography}{87}
\expandafter\ifx\csname natexlab\endcsname\relax\def\natexlab#1{#1}\fi

\bibitem[{{Aoki} {et~al.}(2002){Aoki}, {Ando}, {Honda}, {Iye}, {Izumiura},
  {Kajino}, {Kambe}, {Kawanomonoto}, {Noguchi}, {Okita}, {Sadakane}, {Sato},
  {Shelton}, {Takada-Hidai}, {Takeda}, {Watanabe}, \& {Yoshida}}]{AOK02b}
{Aoki}, W., {Ando}, H., {Honda}, S., {et~al.} 2002, \pasj, 54, 427

\bibitem[{{Aoki} {et~al.}(2006){Aoki}, {Frebel}, {Christlieb}, {Norris},
  {Beers}, {Minezaki}, {Barklem}, {Honda}, {Takada-Hidai}, {Asplund}, {Ryan},
  {Tsangarides}, {Eriksson}, {Steinhauer}, {Deliyannis}, {Nomoto}, {Fujimoto},
  {Ando}, {Yoshii}, \& {Kajino}}]{AOK06b}
{Aoki}, W., {Frebel}, A., {Christlieb}, N., {et~al.} 2006, \apj, 639, 897

\bibitem[{{Aoki} {et~al.}(2005){Aoki}, {Honda}, {Beers}, {Kajino}, {Ando},
  {Norris}, {Ryan}, {Izumiura}, {Sadakane}, \& {Takada-Hidai}}]{AOK05}
{Aoki}, W., {Honda}, S., {Beers}, T.~C., {et~al.} 2005, \apj, 632, 611

\bibitem[{{Aoki} {et~al.}(2007){Aoki}, {Honda}, {Beers}, {Takada-Hidai},
  {Iwamoto}, {Tominaga}, {Umeda}, {Nomoto}, {Norris}, \& {Ryan}}]{AOK07c}
{Aoki}, W., {Honda}, S., {Beers}, T.~C., {et~al.} 2007, \apj, 660, 747

\bibitem[{{Arcones} {et~al.}(2007){Arcones}, {Janka}, \& {Scheck}}]{AJS07}
{Arcones}, A., {Janka}, H.-T., \& {Scheck}, L. 2007, \aap, 467, 1227

\bibitem[{{Arcones} \& {Thielemann}(2013)}]{ArconesThielemann13}
{Arcones}, A. \& {Thielemann}, F.-K. 2013, Journal of Physics G Nuclear
  Physics, 40, 013201

\bibitem[{{Argast} {et~al.}(2004){Argast}, {Samland}, {Thielemann}, \&
  {Qian}}]{Argast04}
{Argast}, D., {Samland}, M., {Thielemann}, F.-K., \& {Qian}, Y.-Z. 2004, \aap,
  416, 997

\bibitem[{{Arlandini} {et~al.}(1999){Arlandini}, {K{\"a}ppeler}, {Wisshak},
  {Gallino}, {Lugaro}, {Busso}, \& {Straniero}}]{Arla99}
{Arlandini}, C., {K{\"a}ppeler}, F., {Wisshak}, K., {et~al.} 1999, \apj, 525,
  886

\bibitem[{{Barbuy} {et~al.}(2014){Barbuy}, {Chiappini}, {Cantelli}, {Depagne},
  {Pignatari}, {Hirschi}, {Cescutti}, {Ortolani}, {Hill}, {Zoccali}, {Minniti},
  {Trevisan}, {Bica}, \& {G{\'o}mez}}]{Barbuy2014}
{Barbuy}, B., {Chiappini}, C., {Cantelli}, E., {et~al.} 2014, \aap, 570, A76

\bibitem[{{Barklem} {et~al.}(2005){Barklem}, {Christlieb}, {Beers}, {Hill},
  {Bessell}, {Holmberg}, {Marsteller}, {Rossi}, {Zickgraf}, \&
  {Reimers}}]{BAR05}
{Barklem}, P.~S., {Christlieb}, N., {Beers}, T.~C., {et~al.} 2005, \aap, 439,
  129

\bibitem[{{Bauswein} {et~al.}(2013){Bauswein}, {Goriely}, \&
  {Janka}}]{Bauswein13}
{Bauswein}, A., {Goriely}, S., \& {Janka}, H.-T. 2013, \apj, 773, 78

\bibitem[{{Belczynski} {et~al.}(2002){Belczynski}, {Kalogera}, \&
  {Bulik}}]{Bel2002}
{Belczynski}, K., {Kalogera}, V., \& {Bulik}, T. 2002, \apj, 572, 407

\bibitem[{{Bonifacio} {et~al.}(2009){Bonifacio}, {Spite}, {Cayrel}, {Hill},
  {Spite}, {Fran{\c c}ois}, {Plez}, {Ludwig}, {Caffau}, {Molaro}, {Depagne},
  {Andersen}, {Barbuy}, {Beers}, {Nordstr{\"o}m}, \& {Primas}}]{BON09}
{Bonifacio}, P., {Spite}, M., {Cayrel}, R., {et~al.} 2009, \aap, 501, 519

\bibitem[{{Caughlan} \& {Fowler}(1988)}]{CF88}
{Caughlan}, G.~R. \& {Fowler}, W.~A. 1988, Atomic Data and Nuclear Data Tables,
  40, 283

\bibitem[{{Cescutti}(2007)}]{Cescutti07}
{Cescutti}, G. 2007, PhD thesis, PhD Thesis, 2007

\bibitem[{{Cescutti}(2008)}]{Cesc08}
{Cescutti}, G. 2008, \aap, 481, 691

\bibitem[{{Cescutti} \& {Chiappini}(2010)}]{Cesc10}
{Cescutti}, G. \& {Chiappini}, C. 2010, \aap, 515, A102

\bibitem[{{Cescutti} \& {Chiappini}(2014)}]{Cescutti14}
{Cescutti}, G. \& {Chiappini}, C. 2014, \aap, 565, A51

\bibitem[{{Cescutti} {et~al.}(2013){Cescutti}, {Chiappini}, {Hirschi},
  {Meynet}, \& {Frischknecht}}]{Cescutti13}
{Cescutti}, G., {Chiappini}, C., {Hirschi}, R., {Meynet}, G., \&
  {Frischknecht}, U. 2013, \aap, 553, A51

\bibitem[{{Cescutti} {et~al.}(2006){Cescutti}, {Fran{\c c}ois}, {Matteucci},
  {Cayrel}, \& {Spite}}]{Cesc06}
{Cescutti}, G., {Fran{\c c}ois}, P., {Matteucci}, F., {Cayrel}, R., \& {Spite},
  M. 2006, \aap, 448, 557

\bibitem[{{Chiappini} {et~al.}(2008){Chiappini}, {Ekstr{\"o}m}, {Meynet},
  {Hirschi}, {Maeder}, \& {Charbonnel}}]{Chiappini08}
{Chiappini}, C., {Ekstr{\"o}m}, S., {Meynet}, G., {et~al.} 2008, \aap, 479, L9

\bibitem[{{Chiappini} {et~al.}(2011){Chiappini}, {Frischknecht}, {Meynet},
  {Hirschi}, {Barbuy}, {Pignatari}, {Decressin}, \& {Maeder}}]{Chiappini11}
{Chiappini}, C., {Frischknecht}, U., {Meynet}, G., {et~al.} 2011, \nat, 472,
  454

\bibitem[{{Chiappini} {et~al.}(2006){Chiappini}, {Hirschi}, {Meynet},
  {Ekstr{\"o}m}, {Maeder}, \& {Matteucci}}]{Chiappini06}
{Chiappini}, C., {Hirschi}, R., {Meynet}, G., {et~al.} 2006, \aap, 449, L27

\bibitem[{{Cohen} {et~al.}(2008){Cohen}, {Christlieb}, {McWilliam}, {Shectman},
  {Thompson}, {Melendez}, {Wisotzki}, \& {Reimers}}]{COH08}
{Cohen}, J.~G., {Christlieb}, N., {McWilliam}, A., {et~al.} 2008, \apj, 672,
  320

\bibitem[{{Cowan} {et~al.}(2002){Cowan}, {Sneden}, {Burles}, {Ivans}, {Beers},
  {Truran}, {Lawler}, {Primas}, {Fuller}, {Pfeiffer}, \& {Kratz}}]{COW02}
{Cowan}, J.~J., {Sneden}, C., {Burles}, S., {et~al.} 2002, \apj, 572, 861

\bibitem[{{Cowan} {et~al.}(1991){Cowan}, {Thielemann}, \& {Truran}}]{Cowan1991}
{Cowan}, J.~J., {Thielemann}, F.-K., \& {Truran}, J.~W. 1991, \physrep, 208,
  267

\bibitem[{{Cristallo} {et~al.}(2011){Cristallo}, {Piersanti}, {Straniero},
  {Gallino}, {Dom{\'{\i}}nguez}, {Abia}, {Di Rico}, {Quintini}, \&
  {Bisterzo}}]{Cristallo11}
{Cristallo}, S., {Piersanti}, L., {Straniero}, O., {et~al.} 2011, \apjs, 197,
  17

\bibitem[{{Cristallo} {et~al.}(2009){Cristallo}, {Straniero}, {Gallino},
  {Piersanti}, {Dom{\'{\i}}nguez}, \& {Lederer}}]{Cristallo09}
{Cristallo}, S., {Straniero}, O., {Gallino}, R., {et~al.} 2009, \apj, 696, 797

\bibitem[{{D'Orazi} {et~al.}(2013){D'Orazi}, {Lugaro}, {Campbell}, {Bragaglia},
  {Carretta}, {Gratton}, {Lucatello}, \& {D'Antona}}]{Dorazi13}
{D'Orazi}, V., {Lugaro}, M., {Campbell}, S.~W., {et~al.} 2013, \apj, 776, 59

\bibitem[{{Fran{\c c}ois} {et~al.}(2007){Fran{\c c}ois}, {Depagne}, {Hill},
  {Spite}, {Spite}, {Plez}, {Beers}, {Andersen}, {James}, {Barbuy}, {Cayrel},
  {Bonifacio}, {Molaro}, {Nordstr{\"o}m}, \& {Primas}}]{Franc07}
{Fran{\c c}ois}, P., {Depagne}, E., {Hill}, V., {et~al.} 2007, \aap, 476, 935

\bibitem[{{Frebel}(2010)}]{Frebel10}
{Frebel}, A. 2010, Astronomische Nachrichten, 331, 474

\bibitem[{{Frebel} {et~al.}(2009){Frebel}, {Johnson}, \& {Bromm}}]{Frebel09}
{Frebel}, A., {Johnson}, J.~L., \& {Bromm}, V. 2009, \mnras, 392, L50

\bibitem[{{Frischknecht} {et~al.}(2012){Frischknecht}, {Hirschi}, \&
  {Thielemann}}]{Frisch12}
{Frischknecht}, U., {Hirschi}, R., \& {Thielemann}, F.-K. 2012, \aap, 538, L2

\bibitem[{{Greggio}(2005)}]{Greggio05}
{Greggio}, L. 2005, \aap, 441, 1055

\bibitem[{{Hayek} {et~al.}(2009){Hayek}, {Wiesendahl}, {Christlieb},
  {Eriksson}, {Korn}, {Barklem}, {Hill}, {Beers}, {Farouqi}, {Pfeiffer}, \&
  {Kratz}}]{HAY09}
{Hayek}, W., {Wiesendahl}, U., {Christlieb}, N., {et~al.} 2009, \aap, 504, 511

\bibitem[{{Honda} {et~al.}(2004){Honda}, {Aoki}, {Kajino}, {Ando}, {Beers},
  {Izumiura}, {Sadakane}, \& {Takada-Hidai}}]{Honda04}
{Honda}, S., {Aoki}, W., {Kajino}, T., {et~al.} 2004, \apj, 607, 474

\bibitem[{{Hotokezaka} {et~al.}(2013){Hotokezaka}, {Kiuchi}, {Kyutoku},
  {Muranushi}, {Sekiguchi}, {Shibata}, \& {Taniguchi}}]{Hoto13}
{Hotokezaka}, K., {Kiuchi}, K., {Kyutoku}, K., {et~al.} 2013, \prd, 88, 044026

\bibitem[{{Ishimaru} \& {Wanajo}(1999)}]{Ishimaru99}
{Ishimaru}, Y. \& {Wanajo}, S. 1999, \apjl, 511, L33

\bibitem[{{Ivans} {et~al.}(2006){Ivans}, {Simmerer}, {Sneden}, {Lawler},
  {Cowan}, {Gallino}, \& {Bisterzo}}]{IVA06}
{Ivans}, I.~I., {Simmerer}, J., {Sneden}, C., {et~al.} 2006, \apj, 645, 613

\bibitem[{{Ivans} {et~al.}(2003){Ivans}, {Sneden}, {James}, {Preston},
  {Fulbright}, {H{\"o}flich}, {Carney}, \& {Wheeler}}]{IVA03}
{Ivans}, I.~I., {Sneden}, C., {James}, C.~R., {et~al.} 2003, \apj, 592, 906

\bibitem[{{Kalogera} {et~al.}(2001){Kalogera}, {Narayan}, {Spergel}, \&
  {Taylor}}]{Kalogera2001}
{Kalogera}, V., {Narayan}, R., {Spergel}, D.~N., \& {Taylor}, J.~H. 2001, \apj,
  556, 340

\bibitem[{{K{\"a}ppeler} {et~al.}(2011){K{\"a}ppeler}, {Gallino}, {Bisterzo},
  \& {Aoki}}]{Kaeppeler11}
{K{\"a}ppeler}, F., {Gallino}, R., {Bisterzo}, S., \& {Aoki}, W. 2011, Reviews
  of Modern Physics, 83, 157

\bibitem[{{Korobkin} {et~al.}(2012){Korobkin}, {Rosswog}, {Arcones}, \&
  {Winteler}}]{Korobkin12}
{Korobkin}, O., {Rosswog}, S., {Arcones}, A., \& {Winteler}, C. 2012, \mnras,
  426, 1940

\bibitem[{{Kyutoku} {et~al.}(2013){Kyutoku}, {Ioka}, \& {Shibata}}]{Kyutoku13}
{Kyutoku}, K., {Ioka}, K., \& {Shibata}, M. 2013, \prd, 88, 041503

\bibitem[{{Lai} {et~al.}(2008){Lai}, {Bolte}, {Johnson}, {Lucatello}, {Heger},
  \& {Woosley}}]{LAI08}
{Lai}, D.~K., {Bolte}, M., {Johnson}, J.~A., {et~al.} 2008, \apj, 681, 1524

\bibitem[{{Lai} {et~al.}(2007){Lai}, {Johnson}, {Bolte}, \&
  {Lucatello}}]{LAI07}
{Lai}, D.~K., {Johnson}, J.~A., {Bolte}, M., \& {Lucatello}, S. 2007, \apj,
  667, 1185

\bibitem[{{Li} {et~al.}(2010){Li}, {Christlieb}, {Sch{\"o}rck}, {Norris},
  {Bessell}, {Yong}, {Beers}, {Lee}, {Frebel}, \& {Zhao}}]{Li10}
{Li}, H.~N., {Christlieb}, N., {Sch{\"o}rck}, T., {et~al.} 2010, \aap, 521, A10

\bibitem[{{Maeder} \& {Meynet}(1989)}]{MM89}
{Maeder}, A. \& {Meynet}, G. 1989, \aap, 210, 155

\bibitem[{{Maeder} {et~al.}(2014){Maeder}, {Meynet}, \& {Chiappini}}]{Maeder14}
{Maeder}, A., {Meynet}, G., \& {Chiappini}, C. 2014, ArXiv e-prints

\bibitem[{{Masseron} {et~al.}(2010){Masseron}, {Johnson}, {Plez}, {van Eck},
  {Primas}, {Goriely}, \& {Jorissen}}]{Masseron10}
{Masseron}, T., {Johnson}, J.~A., {Plez}, B., {et~al.} 2010, \aap, 509, A93

\bibitem[{{Masseron} {et~al.}(2006){Masseron}, {van Eck}, {Famaey}, {Goriely},
  {Plez}, {Siess}, {Beers}, {Primas}, \& {Jorissen}}]{MAS06}
{Masseron}, T., {van Eck}, S., {Famaey}, B., {et~al.} 2006, \aap, 455, 1059

\bibitem[{{Matteucci} {et~al.}(2014){Matteucci}, {Romano}, {Arcones},
  {Korobkin}, \& {Rosswog}}]{Matteucci2014}
{Matteucci}, F., {Romano}, D., {Arcones}, A., {Korobkin}, O., \& {Rosswog}, S.
  2014, \mnras, 438, 2177

\bibitem[{{McWilliam}(1998)}]{MCW98}
{McWilliam}, A. 1998, \aj, 115, 1640

\bibitem[{{McWilliam} {et~al.}(1995){McWilliam}, {Preston}, {Sneden}, \&
  {Searle}}]{MCW95}
{McWilliam}, A., {Preston}, G.~W., {Sneden}, C., \& {Searle}, L. 1995, \aj,
  109, 2757

\bibitem[{{Mennekens} \& {Vanbeveren}(2014)}]{Mennekens14}
{Mennekens}, N. \& {Vanbeveren}, D. 2014, \aap, 564, A134

\bibitem[{{Nakar} {et~al.}(2006){Nakar}, {Gal-Yam}, \& {Fox}}]{Nakar2006}
{Nakar}, E., {Gal-Yam}, A., \& {Fox}, D.~B. 2006, \apj, 650, 281

\bibitem[{{Nishimura} {et~al.}(2015){Nishimura}, {Takiwaki}, \&
  {Thielemann}}]{Nishimura15}
{Nishimura}, N., {Takiwaki}, T., \& {Thielemann}, F.-K. 2015, ArXiv e-prints

\bibitem[{{Oechslin} {et~al.}(2007){Oechslin}, {Janka}, \&
  {Marek}}]{Oechslin07}
{Oechslin}, R., {Janka}, H.-T., \& {Marek}, A. 2007, \aap, 467, 395

\bibitem[{{Pignatari} {et~al.}(2008){Pignatari}, {Gallino}, {Meynet},
  {Hirschi}, {Herwig}, \& {Wiescher}}]{Pigna08}
{Pignatari}, M., {Gallino}, R., {Meynet}, G., {et~al.} 2008, \apjl, 687, L95

\bibitem[{{Piran} \& {Shaviv}(2005)}]{Piran2005}
{Piran}, T. \& {Shaviv}, N.~J. 2005, Physical Review Letters, 94, 051102

\bibitem[{{Preston} {et~al.}(2006){Preston}, {Sneden}, {Thompson}, {Shectman},
  \& {Burley}}]{PRE06}
{Preston}, G.~W., {Sneden}, C., {Thompson}, I.~B., {Shectman}, S.~A., \&
  {Burley}, G.~S. 2006, \aj, 132, 85

\bibitem[{{Roederer} {et~al.}(2014){Roederer}, {Cowan}, {Preston}, {Shectman},
  {Sneden}, \& {Thompson}}]{Roederer14}
{Roederer}, I.~U., {Cowan}, J.~J., {Preston}, G.~W., {et~al.} 2014, \mnras,
  445, 2970

\bibitem[{{Roederer} {et~al.}(2008){Roederer}, {Frebel}, {Shetrone}, {Allende
  Prieto}, {Rhee}, {Gallino}, {Bisterzo}, {Sneden}, {Beers}, \&
  {Cowan}}]{ROE08}
{Roederer}, I.~U., {Frebel}, A., {Shetrone}, M.~D., {et~al.} 2008, \apj, 679,
  1549

\bibitem[{{Romano} {et~al.}(2010){Romano}, {Karakas}, {Tosi}, \&
  {Matteucci}}]{RKT10}
{Romano}, D., {Karakas}, A.~I., {Tosi}, M., \& {Matteucci}, F. 2010, \aap, 522,
  A32+

\bibitem[{{Rosswog}(2013)}]{Rosswog13}
{Rosswog}, S. 2013, Royal Society of London Philosophical Transactions Series
  A, 371, 20272

\bibitem[{{Rosswog} {et~al.}(2000){Rosswog}, {Davies}, {Thielemann}, \&
  {Piran}}]{Rosswog00}
{Rosswog}, S., {Davies}, M.~B., {Thielemann}, F.-K., \& {Piran}, T. 2000, \aap,
  360, 171

\bibitem[{{Rosswog} {et~al.}(1999){Rosswog}, {Liebend{\"o}rfer}, {Thielemann},
  {Davies}, {Benz}, \& {Piran}}]{Rosswog99}
{Rosswog}, S., {Liebend{\"o}rfer}, M., {Thielemann}, F.-K., {et~al.} 1999,
  \aap, 341, 499

\bibitem[{{Scalo}(1986)}]{Scalo86}
{Scalo}, J.~M. 1986, \fcp, 11, 1

\bibitem[{{Shen} {et~al.}(2014){Shen}, {Cooke}, {Ramirez-Ruiz}, {Madau},
  {Mayer}, \& {Guedes}}]{Shen14}
{Shen}, S., {Cooke}, R., {Ramirez-Ruiz}, E., {et~al.} 2014, ArXiv e-prints

\bibitem[{{Simmerer} {et~al.}(2004){Simmerer}, {Sneden}, {Cowan}, {Collier},
  {Woolf}, \& {Lawler}}]{SSC04}
{Simmerer}, J., {Sneden}, C., {Cowan}, J.~J., {et~al.} 2004, \apj, 617, 1091

\bibitem[{{Sneden} {et~al.}(2008){Sneden}, {Cowan}, \& {Gallino}}]{Sneden08}
{Sneden}, C., {Cowan}, J.~J., \& {Gallino}, R. 2008, \araa, 46, 241

\bibitem[{{Thielemann} {et~al.}(2011){Thielemann}, {Arcones}, {K{\"a}ppeli},
  {Liebend{\"o}rfer}, {Rauscher}, {Winteler}, {Fr{\"o}hlich}, {Dillmann},
  {Fischer}, {Martinez-Pinedo}, {Langanke}, {Farouqi}, {Kratz}, {Panov}, \&
  {Korneev}}]{Thielemann11}
{Thielemann}, F.-K., {Arcones}, A., {K{\"a}ppeli}, R., {et~al.} 2011, Progress
  in Particle and Nuclear Physics, 66, 346

\bibitem[{{Thornton} {et~al.}(1998){Thornton}, {Gaudlitz}, {Janka}, \&
  {Steinmetz}}]{Thornton98}
{Thornton}, K., {Gaudlitz}, M., {Janka}, H.-T., \& {Steinmetz}, M. 1998, \apj,
  500, 95

\bibitem[{{Travaglio} {et~al.}(1999){Travaglio}, {Galli}, {Gallino}, {Busso},
  {Ferrini}, \& {Straniero}}]{Trava99}
{Travaglio}, C., {Galli}, D., {Gallino}, R., {et~al.} 1999, \apj, 521, 691

\bibitem[{{Travaglio} {et~al.}(2004){Travaglio}, {Gallino}, {Arnone}, {Cowan},
  {Jordan}, \& {Sneden}}]{Trava04}
{Travaglio}, C., {Gallino}, R., {Arnone}, E., {et~al.} 2004, \apj, 601, 864

\bibitem[{{Truran}(1981)}]{Truran81}
{Truran}, J.~W. 1981, \aap, 97, 391

\bibitem[{{van de Voort} {et~al.}(2015){van de Voort}, {Quataert}, {Hopkins},
  {Kere{\v s}}, \& {Faucher-Gigu{\`e}re}}]{vandenVoort15}
{van de Voort}, F., {Quataert}, E., {Hopkins}, P.~F., {Kere{\v s}}, D., \&
  {Faucher-Gigu{\`e}re}, C.-A. 2015, \mnras, 447, 140

\bibitem[{{Wallerstein} {et~al.}(2007){Wallerstein}, {Kovtyukh}, \&
  {Andrievsky}}]{Wallerstein07}
{Wallerstein}, G., {Kovtyukh}, V.~V., \& {Andrievsky}, S.~M. 2007, \aj, 133,
  1373

\bibitem[{{Wanajo} {et~al.}(2011){Wanajo}, {Janka}, \& {M{\"u}ller}}]{WJM11}
{Wanajo}, S., {Janka}, H.-T., \& {M{\"u}ller}, B. 2011, \apjl, 726, L15

\bibitem[{{Wanajo} {et~al.}(2001){Wanajo}, {Kajino}, {Mathews}, \&
  {Otsuki}}]{Wanajo2001}
{Wanajo}, S., {Kajino}, T., {Mathews}, G.~J., \& {Otsuki}, K. 2001, \apj, 554,
  578

\bibitem[{{Westin} {et~al.}(2000){Westin}, {Sneden}, {Gustafsson}, \&
  {Cowan}}]{WES00}
{Westin}, J., {Sneden}, C., {Gustafsson}, B., \& {Cowan}, J.~J. 2000, \apj,
  530, 783

\bibitem[{{Winteler} {et~al.}(2012){Winteler}, {K{\"a}ppeli}, {Perego},
  {Arcones}, {Vasset}, {Nishimura}, {Liebend{\"o}rfer}, \&
  {Thielemann}}]{Winteler12}
{Winteler}, C., {K{\"a}ppeli}, R., {Perego}, A., {et~al.} 2012, \apjl, 750, L22

\bibitem[{{Woosley} {et~al.}(1994){Woosley}, {Wilson}, {Mathews}, {Hoffman}, \&
  {Meyer}}]{WWM94}
{Woosley}, S.~E., {Wilson}, J.~R., {Mathews}, G.~J., {Hoffman}, R.~D., \&
  {Meyer}, B.~S. 1994, \apj, 433, 229

\bibitem[{{Yong} {et~al.}(2014){Yong}, {Alves Brito}, {Da Costa},
  {Alonso-Garc{\'{\i}}a}, {Karakas}, {Pignatari}, {Roederer}, {Aoki},
  {Fishlock}, {Grundahl}, \& {Norris}}]{Yong14}
{Yong}, D., {Alves Brito}, A., {Da Costa}, G.~S., {et~al.} 2014, \mnras, 439,
  2638

\bibitem[{{Yong} {et~al.}(2006){Yong}, {Aoki}, {Lambert}, \&
  {Paulson}}]{Yong06}
{Yong}, D., {Aoki}, W., {Lambert}, D.~L., \& {Paulson}, D.~B. 2006, \apj, 639,
  918

\bibitem[{{Yong} {et~al.}(2008){Yong}, {Lambert}, {Paulson}, \&
  {Carney}}]{Yong08}
{Yong}, D., {Lambert}, D.~L., {Paulson}, D.~B., \& {Carney}, B.~W. 2008, \apj,
  673, 854

\bibitem[{{Yoon} {et~al.}(2006){Yoon}, {Langer}, \& {Norman}}]{Yoon06}
{Yoon}, S.-C., {Langer}, N., \& {Norman}, C. 2006, \aap, 460, 199

\end{thebibliography}
\end{document}